# Normalisation de la langue et de l'écriture arabe : enjeux culturels régionaux et mondiaux

**Henri Hudrisier**
Laboratoire Paragraphe Université Paris 8,
Liaison de l'AUF à l'ISO SC36
Henri.hudrisier@wanadoo.fr

**Mokhtar Ben Henda**
Laboratoire MICA, Université Bordeaux 3
*Convener* du SC36/WG1
mbenhenda@u-bordeaux3.fr

**Résumé** :

La langue et l'écriture arabe sont aujourd'hui confrontées à une recrudescence de solutions normatives internationales qui remettent en cause la plupart de leurs principes de fonctionnement en site ou sur les réseaux. Même si les solutions du codage numérique multilingue, notamment celles proposées par Unicode, ont résolu beaucoup de difficultés de l'écriture arabe, le volet linguistique est encore en quête de solutions plus adaptées. La terminologie est l'un des secteurs dans lequel la langue arabe nécessite une modernisation profonde de ses modèles classiques de production. La voie normative, notamment celle du TC37 de l'ISO, est proposée comme une des solutions qui lui permettrait de se mettre en synergie avec les référentiels internationaux pour mieux intégrer la société du savoir en voie de construction.

**Mots clés** : Langue arabe, normalisation, terminologie, multilinguisme

**Abstract:**

Arabic language and writing are now facing a resurgence of international normative solutions that challenge most of their local or network based operating principles. Even if the multilingual digital coding solutions, especially those proposed by Unicode, have solved many difficulties of Arabic writing, the linguistic aspect is still in search of more adapted solutions. Terminology is one of the sectors in which the Arabic language requires a deep modernization of its classical productivity models. The normative approach, in particular that of the ISO TC37, is proposed as one of the solutions that would allow it to combine with international standards to better integrate the knowledge society under construction.

## 1. Introduction

Toutes les langues et toutes les écritures du monde sont actuellement confrontées à des enjeux normatifs urgents et stratégiques qui mettent en question le devenir communicationnel, cognitif et même économique des communautés concernées. C'est en se limitant à cette facette d'approche qu'un chercheur non arabophone peut utilement contribuer à un groupe de travail sur les enjeux numériques de la langue et de l'écriture arabe. Mais la synergie des deux profils linguistiques et des expériences des deux auteurs de ce papier nous permet de dépasser ce premier niveau d'analyse. Nous explorerons les soubassements culturels et techniques de l'évolution de la langue arabe d'un point de vue normatif, mais aussi son adéquation avec les innovations de la technologie numérique dans le traitement multilingue de l'information et de la communication.

## 2. Rappel méthodologique, rappel historique

Les langues et les écritures entretiennent des rapports étroits, mais il est important de ne pas confondre l'un avec l'autre. Les langues préexistent indépendamment des écritures mais dès lors que la communication écrite est adoptée par une communauté linguistique, il se développe un (ou plusieurs) système(s) d'écriture spécifiquement inventé(s) par la communauté linguistique en question, ou emprunté(s) à une autre tradition linguistique. Cette notion est évidemment toute relative. On peut par exemple considérer avec raison, que l'écriture arabe a été spécifiquement inventée pour transcrire la langue arabe dans sa phase d'évolution du proto arabe pré coranique (nabatéen)[1] jusqu'à un état considéré comme définitif (arabe littéraire ou coranique). D'évidence cette écriture est (ou a été) empruntée pour transcrire de nombreuses autres langues avec ou sans ajouts mineurs sur l'alphabet arabe standard (ourdou, turc ottoman, berbère, de nombreuses langues africaines, perses, langues de l'ex-union soviétique)[2]. Tel est l'état de spécificité scripto-linguistique que l'ont peut constater dans d'autre situations similaires qui constituent de façon « fondatrice », pourrait-on dire, les différentes grandes familles d'écritures du monde : latine, cyrillique, indienne[3], chinoise. L'écriture arabe appartient à ce « top cinq[4] » des grandes écritures du monde qui la met dans une situation intéressante qui se doit d'être préservée. Elle constitue de fait, un des registres majeurs des normes des écritures du monde. Certes, pour des raisons historiques et de simplicité formelle (mais aussi démographico-linguistique) l'écriture latine non accentuée bénéficie d'une position pivot et dominante dans ces 5 familles d'écritures[5] ; mais cette domination de fait de l'écriture latine[6] est actuellement disputée par le catalogue universel de toutes les écritures du monde réunies dans Unicode. Pour des raisons évidentes de supériorité industrielle, démographique et économique, l'Asie du Sud et de l'Est prennent une place telle dans le développement des TIC que leurs écritures existent aujourd'hui dans un contexte d'ingénierie numérique et d'interopérabilité normative remarquable. L'écriture arabe, quant à elle, ne s'est pas entièrement affranchie de son legs culturel et historique d'écriture sacrée pour affronter les défis de l'industrialisation et de la numérisation. Rappelons ici que jusqu'au XXe siècle, l'imprimerie arabe est restée confinée à un rôle marginal, nullement comparable à l'essor fantastique de son homologue européenne. Son influence est restée limitée, les livres imprimés sur les presses arabophones de l'époque constituent des trésors de la typographie arabe, mais ils sont difficilement reproduits et peu diffusés du fait des difficultés techniques que leur impression engendrait.

---

[1] La descendance de l'arabe de l'alphabet nabatéen, et par lui, de l'araméen est définitivement prouvée. Le nabatéen (IIe siècle av. JC), créé chez une tribu arabe de la mer Morte (Petra et Bostra) pour écrire l'araméen, se distingua par la réunion des lettres les unes aux autres au moyen de ligatures. Cette caractéristique est à la base de l'écriture arabe contemporaine. C'est plutôt la transformation de l'araméen à l'alphabet arabe moderne qui reste encore incertaine puisque l'alphabet arabe n'est attesté qu'à partir du VIe siècle ap. JC, un siècle avant l'arrivée de l'Islam qui permis sa diffusion large dans les pays conquis.

[2] Le persan, l'afghan et certains parlers d'Afrique. Quant aux Turcs ce n'est que récemment qu'ils optèrent pour l'alphabet latin au détriment de l'arabe. La situation berbère est plus complexe. On ne peut nier l'arrivée très ancienne du libyque sur le territoire africain, la permanence de l'écriture tifinagh chez les Touaregs, l'adoption de l'écriture arabe (notamment dans le Mzab), l'influence de la colonisation (mais aussi des linguistes et de certains missionnaires) pour la notation des langues berbères en caractères latins avec ou sans ajouts diacritiques et dernièrement la volonté culturelle identitaire pour se réapproprier une écriture plusieurs fois millénaire (au Maroc mais aussi en Kabylie).

[3] Qui est un cas particulier du fait de la variation formelle des écritures apparentées à la famille dite des écritures indiennes (indi mais aussi thaï, cambodgien, laotien, malaisien, etc…)

[4] La plupart des autres écritures sont très spécifiques à une langue (coréen, hébraïque, arménien, …) ou, comme le japonais, elles sont une adaptation mixte (idéographie chinoise [kanji] & syllabaires kana).

[5] C'est dans l'espace alphabétique latin qu'a été « réinventée » l'imprimerie qui a permis le développement scientifique, industriel et mondialisé que nous connaissons aujourd'hui. C'est aussi dans cet alphabet latin que s'est développé la télégraphie dont découlent les média électronique et numérique actuel.

[6] Caractère isolés, unicité des formes.

Sur le plan numérique, alors que les cultures asiatiques se sont alliées – malgré leurs divergences linguistiques – autour d'une norme CJK (coréens, japonais et chinois)[7] pour contrer l'hégémonie technologique latine du codage informatique à 7 et 8 bits et produire un codage à 16 bits plus adapté à la transcription de leurs écritures, les éditeurs arabes n'ont pas réussi, malgré leur unicité culturelle et linguistique et malgré les initiatives de l'ASMO[8] et des versions multiples de la norme CODAR, à fonder par eux-mêmes une norme unifiée de codage de l'écriture arabe. Il a fallu l'intervention des industriels occidentaux comme l'ECMA[9] pour fixer et industrialiser les premières normes de codification de la langue arabe. Les initiatives de normalisation internationales, particulièrement les normes ISO 8859, Unicode et ISO 10646, ont fini par stabiliser (relativement[10]) la forme de codage informatique de l'écriture arabe et à l'intégrer (quoique comme langue encore minoritaire) sur Internet[11]. Un rappel des acteurs clés d'une stratégie arabe de normalisation aiderait à mieux poser les défis d'avenir, surtout en rapport avec la normalisation des contenus par l'harmonisation d'une terminologie et d'une sémantique arabe unifiée.

## 3. La normalisation technologique arabe : enjeux et défis

Avant de parler de techniques relatives à l'harmonisation terminologique et sémantique de la langue arabe, il est primordial d'évoquer des questions d'ordre stratégique quant aux modalités opératoires des instances régionales de la normalisation dans l'appui à la présence de la langue arabe dans les circuits et les cercles de décision internationales. Á l'ère de la mondialisation, la normalisation constitue en effet, un passage obligé dans le développement des sciences et des techniques pour toutes les nations. Dans le monde arabe, étant donné que le secteur des TIC connaît une croissance remarquable et constitue graduellement un élément essentiel de son économie, le développement de normes arabes pour les TIC est de plus en plus impératif. Les normes ont prouvé leur capacité à promouvoir des activités d'e-commerce, e-gouvernement, e-contenu et e-Learning. Elles sont donc en mesure de contribuer aux transformations économiques et sociales de la région malgré les défis constants qui ont souvent ralenti leur application.

Parmi les grands défis à la langue arabe à l'ère du numérique, il y avait d'abord le codage des caractères qui a été largement résolu par les contributions endogènes et exogènes de l'ASMO, l'ECMA, le consortium Unicode et l'ISO. Il y avait ensuite (les années 1990-2000) les défis de l'internationalisation (i18n) et de la localisation (l10n) des logiciels et des contenus, deux procédés permettant de rendre accessibles des contenus Internet au plus grand nombre d'usagers à travers la planète et aussi d'adapter ces mêmes contenus à des environnements à fort caractère local. Aujourd'hui, la localisation et l'internationalisation de logiciels et de contenus sont les deux formes d'un seul processus de diffusion ou de commercialisation internationale.

---

[7] CJK est un groupe de langues asiatiques dont l'écriture habituelle ou traditionnelle utilise les sinogrammes. On rencontre aussi l'abréviation CJKV, ajoutant le vietnamien, qui s'écrivait aussi en sinogrammes.

[8] ASMO : *Arab Standards and Metrology Organization*

[9] Le *European Computer Manufactures Association*, autorité d'enregistrement de l'ISO pour les codes informatiques, a proposé en février 1982 un code pléthorique destiné à satisfaire les attentes des partenaires arabes en discussions inabouties depuis 1976 pour une norme de codage unifié (CODAR1, CODAR2, CODAR-U).

[10] Si l'on considère les problèmes d'interfaces dans les applications informatiques multilingues arabes (i.e. la bidirectionnalité, la localisation, les ligatures, les césures dans les textes multilingues), il est évident que des problèmes de codage de la langue arabe sont encore à résoudre.

[11] Selon le rapport mondial publié par l'Union Internationale des Télécommunications en 2011, le volume estimé de contenu arabe sur l'Internet est d'environ 3%.

Or, aujourd'hui encore, le défi de l'internationalisation et de la localisation des noms de domaines pose des enjeux importants à la présence arabe sur Internet. Le mouvement iDN (*International Domain Name*), gTLD (*general Top Level Domain*) et ccTLD (*country code Top Level Domain*) sont au cœur du processus de la diversité linguistique des réseaux d'information dans le monde. Le développement des noms de domaine multilingues est fondé sur la nécessité d'éliminer la barrière linguistique qui empêche la population ne parlant pas l'anglais d'utiliser Internet. Ces types d'initiatives permettent aux communautés non anglophones de l'Internet d'éviter les problèmes liés à la translittération.

Le mouvement iDN signifie concrètement que les noms de domaine (mais également les extensions) de niveau générique (gTLD) ou spécifiques (ccTLD) du genre « www.domaine.extension » contiennent au moins un caractère régional non ASCII. De plus en plus de noms de domaines contiennent des caractères accentués ou des diacritiques non-latins. Les extensions elles-mêmes existent aujourd'hui en caractères cyrilliques, arabes ou chinois. Il est prévu que dans quelques années, les noms de domaine dans des langues et des caractères locaux deviendront un standard de l'Internet. Or, bien que les ccTLD en caractères latins dans les pays arabes existent depuis longtemps, ils sont encore loin d'occuper une place prépondérante dans le classement mondial. D'après le classement de l'UIT du mois de mai 2011, le premier ccTLD latin d'un pays arabe, celui du Maroc (.ma), arrive en 66$^e$ position devant l'Arabie Saoudite (.sa : 71$^e$) et plus loin la Libye (.ly), la Jordanie (.jo) et la Tunisie (.tn) respectivement aux 79$^e$, 80$^e$ et 81$^e$ rangs. Avec l'iDN (utilisation de caractères locaux dans le nom de domaine), le bilan est encore moins brillant. Le principe est certes nouveau[12] et constitue une révolution en soi, mais sa promotion dans le monde arabe est encore freinée par le faible volume de développements de contenus et le faible niveau de propagande auprès des acteurs économiques et industriels. On a vu ces derniers mois (2012) des projets d'internationalisation de ccTLD qui s'accélèrent et qui sont en passe d'être tous validés. Parmi les extensions internationalisées, celui de l'Arabie saoudite (السعودية.) arrive en 91$^e$ position du classement de 2011, avec 1 708 noms de domaines enregistrés en caractères arabes. D'autres pays comme l'Égypte, la Jordanie, la Palestine, le Qatar, la Tunisie, les Émirats arabes unies, l'Algérie, le Maroc, Oman, la Syrie, ont déposé leurs propositions d'iDN ccTLD dont beaucoup ont été validés par l'ICANN (*Internet Corporation for Assigned Names and Numbers*).

Les pays arabes sont certes loin des modèles chinois ou russes sur ce plan, mais l'on peut aisément se dire que si tous les projets ont été validés aussi vite, c'est que la demande était forte et que les opportunités sont nombreuses. Des améliorations sont toutefois encore à étudier par rapport aux caractéristiques de nommage en langue arabe comme la préférence pour un seul mot dans le choix de nom de domaine, l'usage des acronymes et l'harmonisation des combinaisons de plusieurs mots face à la complexité de l'écriture arabe elle-même. Les extensions des noms de domaines gagneraient aussi à se faire selon une normalisation internationale de noms de pays conforme à la norme ISO 3166, et non dans une forme développée de noms de pays.

L'écriture arabe connaît aussi les problèmes de la reconnaissance optique de caractères dus essentiellement aux problèmes de la mécanisation de l'écriture et de son aspect cursif. La langue arabe représente, en effet, un important facteur d'agglutination qui pose de nombreux défis de reconnaissance de formes et de traitement typographique. Il y a tant de signes dans une écriture arabe voyellée ou dans un style typographique basé sur une ligature superposée, qu'un système d'OCR peine souvent à distinguer les formes des caractères. Souvent, le fait de ressaisir un texte et de le corriger manuellement est plus rentable en temps et en coûts que de le travailler avec un logiciel d'OCR. Pour les documents anciens, comme l'industrie éditoriale était relativement pauvre, le papier et l'encre étaient d'une qualité moyenne qui rend inappropriée une reconnaissance optique optimale de caractères. La combinaison de tous ces facteurs fait qu'une opération d'OCR devient très souvent extrêmement complexe et qu'il est très hasardeux d'en comparer la rentabilité par rapport au seul recours à des opérateurs humains même pour la numérisation de corpus très importants.

---

[12] Les premières demandes d'internationalisation de nom de domaines ont été enregistrées par l'ICANN (*Internet Corporation for Assigned Names and Numbers*) en 2009.

Sur un autre plan relatif aux problèmes des contenus, la société du savoir en construction et les réseaux de la connaissance partagée confrontent l'industrie de la langue arabe aux nouveaux enjeux de la sémantique sur les réseaux numériques. Il ne s'agit plus désormais d'une question de transcription et de codage d'un système d'écriture, mais plus encore d'analyse morphosyntaxique, de traduction automatique, d'intelligence systémique, dans lesquels la terminologie et la taxonomie arabes sont confrontées à un souci majeur d'harmonisation de concepts et d'unification de procédés de développement ontologique.

## 4. Acteurs institutionnels de la normalisation technologique arabe

Tous ces problèmes et tant d'autres que nous n'avons pas cités ici, sont, à notre sens, inhérents à deux préjudices fondamentaux : une faible coordination institutionnelle régionale (et donc un manque de cohérence stratégique entre les pays) et un retard caractérisé dans le patrimoine normatif arabe (cf. annexe). Regardons toutefois comment le monde arabe a réagi sur le plan institutionnel pour faire face à ces défis normatifs ?

D'abord, il faut signaler, comme le remarque Rachid Zghibi, que « jusqu'en 1976, il n'existait aucun code normalisé de transmission de données en caractères arabes qui comportait l'alphabet arabe complet, signes de vocalisation inclus » (Zghibi, 2002). C'est sur initiative conjointe de constructeurs européens (i.e. ECMA) et de représentants arabes que les premières formes de normalisation du caractère arabe ont pris forme avec l'appui du comité technique n°8 « caractères arabes en informatique » créé en 1981 par l'ASMO. Ces initiatives ont abouti à des premiers résultats importants comme dont la norme ASMO 449, équivalente de la norme internationale ISO 903613 pour un codage de caractères sur sept bits. Cette norme a été suivie, trois ans plus tard, d'une tentative de conception d'un ordinateur qui fonctionne uniquement en arabe grâce à une norme de codage sur 8 bits référencée ASMO 662. Le projet n'ayant pas eu de succès, le CT8 de l'ASMO a mis en place en 1988 la norme ASMO 708 pour un bilinguisme arabe/latin (alphabet latin/arabe). ASMO 708 a été homologuée ensuite par l'ISO sous la référence ISO 8859-6. Vers la fin des années 1980, plus de 20 jeux de caractères codés ont été développés pour l'écriture arabe par les industriels informatiques dont les plus connues sont MS Windows Arabic Code Page 1256, Arabic Mac Code Page, Arabic Windows 3X Code Page, Code Page 864 Dos Arabic.

En 1989, l'ASMO a été dissoute et ses fonctions ont été transférées au Centre de normalisation et de métrologie dans le cadre de l'Organisation arabe pour le développement industriel et minier (CSM/AIDMO), spécialisée dans les domaines de l'industrie, des mines et de la normalisation, opérant, depuis sa création en mai 1968, sous l'égide de la Ligue des États arabes. En 1998, le TC-8 de l'ASMO, rebaptisé « Comité technique sur l'utilisation de l'arabe dans les TI » a été à son tour transféré à l'Organisation syrienne arabe de normalisation et de métrologie (SASMO) à Damas.

---

[13] ISO, ISO 9036: information processing-Arabic 7 bits coded character set for information interchange, Geneva, ISO, 1987.

Les initiatives institutionnelles dans le monde arabe ont également touché les autres aspects de la normalisation linguistique sur Internet comme l'arabisation des noms de domaines. L'ICANN (*Internet Corporation for Assigned Names and Numbers*), qui est en partie responsable de la réglementation du réseau Internet, qui contrôle les noms de domaine et qui accorde également les licences aux entreprises qui enregistrent des noms de domaine, a fait des efforts importants pour améliorer la diversité linguistique de l'Internet. Deux organisations internationales sans but lucratif font un travail de pionnier dans ce sens :

- MINC (*Multilingual Internet Names Consortium*) est une ONG internationale sans but lucratif qui promeut à la fois (a) le multilinguisme des noms de domaine ; (b) l'internationalisation des normes et des protocoles des noms de domaines Internet ; (c) la coordination technique avec les organismes internationaux ;

- AINC (*Arabic Internet Names Consortium*) est une association fondée en Avril 2001 pour coordonner les efforts entre les communautés arabophones, afin de promouvoir la croissance du contenu arabe en ligne et de faciliter le développement de nouvelles technologies de l'information au profit de la culture, de la langue, de l'éducation et du développement des compétences arabes, avec une attention particulière à l'arabisation des noms de domaine.

La création de noms de domaine en langue arabe (ADN : *Arabic Domain Name*) implique la nécessité de résoudre des problèmes relatifs aux questions linguistiques, aux noms de domaines arabes de premier niveau (TLD : *Top Level Domain*), aux solutions techniques et aux serveurs racines (*root servers*). Plusieurs solutions ont été développées pour gérer l'arabisation des ADN, mais beaucoup de problèmes persistent encore :

- l'incompatibilité fréquente des solutions proposées ;
- les enregistrements multiples de propositions d'ADN ;
- la non reconnaissance des ADN par l'ICANN ;
- l'isolement des réseaux arabes.

Par conséquent, les demandes fréquentes d'ADNs autonomes a conduit à la mise en place d'un certain nombre d'organisations, dont le Consortium des Noms Internet Arabe (AINC), qui comprend un Comité arabe linguistique (ALC) qui définit des lignes directrices linguistiques. L'ALC a pour but de se charger des actions suivantes :

- Définir le jeu de caractères arabes accepté pour les noms de domaines arabes (ADN) ;
- Définir les TLD de l'arborescence des ADN, à savoir, le niveau générique des noms de domaines arabes (gTLD) et les niveaux des noms de domaines nationaux (ccTLD).

Les préoccupations majeures de l'ALC est de faire en sorte que les noms de domaines arabes préservent les caractéristiques uniques de la langue et de la culture et qu'ils encouragent les locuteurs de la langue arabe à accéder à des sites dans leurs propres langues. Cela suppose des défis culturels et linguistiques majeurs qui, en plus de la diversité des systèmes d'écriture internes au monde arabe (berbère, copte, kurde), relèvent aussi des choix culturels dans l'harmonisation de la langue arabe elle-même dans le domaine terminologique et lexical. Des disparités taxonomiques et sémantiques sont encore très courantes entre les régions (Machreq/Maghreb), dans les pays (villes et campagnes) et dans les structures (langue littéraire ou administrative/langue dialectale ou populaire). Faisons-en ci-après un condensé rapide.

## 5. Stratégies multilinguistiques arabes : atouts multilingues et multiécritures

Depuis une douzaine de siècles le Monde arabe a la chance de partager en commun une langue sacrée, savante, technique, administrative, médiatique… qui transcende ses disparités de cultures linguistiques dialectales. Une langue de plus, qui par sa notation canonique non accentuée est beaucoup moins dépendante d'une improbable dérive phonétique que la plupart des autres langues (européennes notamment). L'arabe classique, est ainsi une langue prestigieuse associée à la religion et à l'écrit, c'est-à-dire à la culture littéraire, à la science et à la technologie et aux fonctions administratives. Elle est aussi appelée arabe coranique, arabe moderne standard, arabe grammatical ou arabe éloquent. Ces caractéristiques ne l'ont pourtant pas préservé des critiques, souvent fondées, quand à sa modernisation lente sur le plan des terminologies techniques. On attribue même à son statut canonique l'inconvénient d'approfondir les écarts entre les dialectes nationaux évolutifs (par emprunts, composition, translitération) et « l'approvisionnement » terminologique dans les langues latines pour combler les vides terminologiques et conceptuels issus du développement socioculturel et technologique[14].

On évoque même dans le statut conventionnel canonique de la langue arabe, des disparités conceptuelles et terminologiques régionales entre *Maghreb* et *Machreq,* inhérentes à l'inertie des académies[15] de la langue arabe dans la normalisation et l'unification des termes et concepts parfois dans des domaines littéraires. Un concept comme « l'encadrement » (i.e. d'une recherche), d'une influence francophone directe sur l'arabe littéraire au Maghreb, donne lieu au terme arabe « Ta'tir : تأطير », entièrement incompréhensible au Moyen-Orient sous influence anglo-saxonne dominante dans laquelle le même concept est traduit par le terme « Ichraf : إشراف » équivalent du terme anglais *supervising*. Les exemples de ce genre sont multiples et peuvent parfois être choquants pour des communautés de même appartenance culturelle[16].

De plus, la quasi-totalité des locuteurs arabes sont au minimum bilingues, bi-écriture quand ce n'est pas plus. Encore faut-il qu'ils ne minorent pas leur langue maternelle (dialectale) et qu'ils apprennent l'arabe classique international. En dehors du français et de l'anglais, ils auraient aussi intérêt à développer l'apprentissage d'autres langues, voire écritures de la zone linguistique arabe (berbère, copte, kurde), ce que beaucoup de palestiniens ont réussi à faire en maîtrisant la langue et l'écriture hébraïque. Bien que ce dernier cas soit une exception par rapport aux autres langues de la région arabe, il traduit à notre sens une dialectique de rapports de forces (politiques mais aussi interculturelles) qu'on observe également dans les rapports des Arabes avec les langues occidentales. Rares sont les Arabes qui apprennent une langue berbère, le kurde ou le copte, alors que l'inverse est très courant voire quasi-indispensable. En revanche, la majorité des Arabes sont francophones ou anglophones alors qu'en proportion, rare sont les Occidentaux qui parlent l'arabe. Cette asymétrie langagière est « naturelle » si on l'explique par la logique mondiale de la hiérarchie des cultures dominantes. Or, en dehors des considérations géopolitiques et des sensibilités des identités culturelles et linguistiques, il y a là un argument économique considérable à ne pas négliger : maitriser 3 langues et 3 écritures est un fabuleux avantage pour apprendre d'autres langues, et notamment pour apprendre les langues non européennes qui deviennent de plus en plus indispensables pour être un acteur efficient dans la mondialisation (Hudrisier, Ben Henda, 2009).

---

[14] À un moment de son histoire, la langue arabe servait de source d'emprunt terminologique et conceptuel à d'autres langues latines. En espagnol, on recense dans les 4 000 mots d'origine arabe. On les identifie souvent par la préposition « Al » au début du mot (almacen, algeciras, alferez). En français on estime ce nombre à 250, empruntés directement ou par l'intermédiaire de l'espagnol et de l'italien (almanaque, chiffre, douane, épinard, estragon, quintal, tasse, chameau).

[15] La multiplicité des académies de la langue arabe dans chacun des pays arabes est en elle-même une source de disparité et de divergence dans la définition de la terminologie arabe.

[16] En arabe littéraire, le terme « Intisab » est utilisé au Maghreb pour signifier l'action de « s'installer » alors qu'au Machreq, il est compris comme l'équivalent du concept « érection ». En Tunisie, le mot « Tabouna » qualifiant le pain de four fait maison, est un mot choquant au Maroc, idem pour le terme « dibara » dans un arabe maghrébin qui veut dire « donner conseil », alors que dans certains pays arabes du Machreq (par exemple en Syrie), il est compris comme un dérivé du substantif « dobr » signifiant le « derrière (le bas-du dos) ».

Une de nos hypothèses serait aussi que le monde arabe, de par sa langue écrite unique, est héritier d'une tradition linguistique particulièrement riche et intéressante qui nécessite une mise en cohérence d'une nomenclature normalisée de concepts. On a tendance dans le monde arabe à glorifier la langue en l'associant au Coran ou en énumérant sa richesse synonymique. Il y aurait selon certaines sources 80 termes différents pour exprimer le miel, 200 pour le serpent, 500 pour le lion, 1000 pour le chameau, autant pour l'épée, et jusqu'à 4000 pour rendre l'idée de malheur. Toute culture est confrontée à une foule de nuances d'idées[17], mais toutes les cultures n'ont pas bénéficié très tôt dans l'histoire, comme la civilisation arabe, des siècles d'or (8ᵉ à 10ᵉ Sc.) et des prestigieux grammairiens et philologues de la cour de Bagdad. Ils ont su théoriser toutes ces subtilités, capter et fixer les termes spécifiques, ce qui fait que, dans le grand nombre d'expressions employées pour une même idée, il y a une foule de figures et de tropes. Un grammairien arabe dit qu'il faudrait 6 chameaux pour transporter le recueil des racines de la langue; un autre auteur prétend avoir compté 12 305 052 mots, en prenant sans doute pour des mots différents les modifications que subit une même racine selon les cas, les nombres, les personnes, les temps, les modes, etc. Il est certain que les racines arabes sont au nombre de 6 000 environ, et que le vocabulaire comprend 60 000 mots (Jorda, 2008). Mais on a tendance à oublier en revanche que le facteur de sacralité religieuse est aussi un facteur d'inertie linguistique et que le facteur de la richesse synonymique est un facteur de confusion et d'incompréhension d'interlocuteurs parlant la même langue[18].

Il est de ce fait indispensable de contribuer pour que les avancées techniques et savantes de la terminologie et de la terminotique (l'algorithme est aussi un apport culturel du monde arabe savant) soient accompagnées par le développement de contenus terminologiques arabes conséquents et cohérents, construits selon des modèles normalisés qui abordent avec beaucoup d'efficacité les questions des nuances conceptuelles et des terminologies appropriées et interopérables.

## 6. La normalisation de la terminologie et de la lexicographie : un enjeu fondamental pour la langue arabe

La terminologie est l'étude des termes qui se réfèrent délibérément à des concepts spécifiques dans des domaines particuliers. Autrement dit, les termes sont toujours étudiés en relation avec le système conceptuel auquel ils appartiennent et dans lequel ils fonctionnent comme dépositaires de la connaissance. Or, on sait bien que l'une des critiques fondamentales fait à la langue arabe est précisément son pauvre taux de néologismes dans les domaines des sciences et des techniques modernes. Par néologismes, on entend les nouvelles unités lexicales ou des unités lexicales existantes qui acquièrent un nouveau sens. En outre, trouver des équivalents en arabe à des termes techniques anglais ou français, engendre de nombreux problèmes en raison de la nature différente de ces langues. Ces problèmes se manifestent notamment du fait des innovations dans domaine des sciences et des technologies ainsi que sous l'effet des médias qui diffusent presque tous les jours, de nouvelles idées et de nouveaux concepts. Tout cela apparaît clairement dans une étude récente d'Ahmed Ramadan El Mgrab de l'Université de Benghazi (2011) portant sur les fondements sociolinguistiques et les aspects lacunaires de la terminologie arabe à l'ère du numérique dont nous ci-après les axes d'analyse.

---

[17] Les Esquimaux avec la neige et la glace, les Indiens et les Mongols avec les chevaux, etc…

[18] Il n'est pas évident que quelqu'un d'une communauté qui utilise le verbe « أنصت » ou « استمع » pour exprimer l'idée d'écouter ou entendre puisse être facilement compris par une communauté employant pour le même concept le verbe « أصغى ». Ces cas de figures sont nombreux en raison d'une synonymie très riche pour l'arabe classique et en raison d'emprunts d'autres langues pour le dialectal (le chat = Herr, Guitt ; Gattous ; Bessa ; Bazzouna).

On peut d'emblée attribuer ces aspects lacunaires aux méthodes anciennes et mais aussi actuelles de constitution des corpus terminologiques par les linguistes terminologues arabes ou à travers les formes d'usage et d'appropriations langagières transmises et héritées d'une génération à l'autre.

Trois méthodes essentielles contribuent depuis le VIe siècle au développement de la terminologie arabe :

- **La dérivation** (إشتقاق) : c'est le fait de dériver la racine d'un terme pour créer un autre, c'est-à-dire, créer de nouveaux termes à partir de racines de mots (radicaux). La dérivation des racines a toujours été considérée depuis la dynastie des Abbasides (VIe s.) comme le moyen le plus naturel d'accroître le corpus terminologique arabe. Dans ce processus, les consonnes radicales ne sont pas modifiées, mais augmentées dans des formes grammaticales multiples (تفعيلات) pour créer des sens dérivés : « طلب/طالب/مطلوب/مطلب/ » = « demander/demandeur/demandé/demande ». La façon la plus simple de générer des termes par dérivation consiste à leur faire partager la même racine verbale trilatérale (radical), sans variation des autres lettres qui donnent le genre morphémique à la dérivation.

  La circonlocution ou la dérivation par traduction (الاشتقاق بالترجمة) est une variante de la méthode de dérivation. Elle consiste à introduire de nouveaux termes en leur donnant la signification des termes étrangers. Il s'agit d'un phénomène universel dans les langues naturelles et non exclusif à la langue arabe. Son usage intensif dans la terminologie arabe est dû à l'abondance de néologismes d'origine étrangère auxquels il n'est pas possible de trouver des équivalents unitermes. En effet, dans plusieurs cas, un mot anglais ou français est traduit vers l'arabe sous forme d'une expression entière comme « Microphone » (جهاز تكبير الصوت) [littéralement : appareil d'amplification de la voix], ou « idéal » (مثل أعلى) [littéralement : exemple supérieur]. Toutefois, la circonlocution, contrairement à d'autres méthodes, semble peu adéquate et conduit à une multiplicité de termes. Elle produit des termes plus longs que le terme d'origine, souvent sous forme de phrases entières qui posent des problèmes syntaxiques. Par conséquent, les traductions de ces néologismes s'écartent souvent de leurs sens fonctionnels réels. Certains traducteurs inventent même leurs propres traductions arbitraires qui ne sont pas conformes aux canons grammaticaux de l'arabe.

- **L'arabisation** (تعريب) est, par définition, l'adaptation des termes non-arabes à la langue arabe en appliquant aux termes étrangers les règles des systèmes phonologiques et parfois morphologiques de la langue comme dans les exemples de : « Philosophie » → (فلسفة = falsafah),

« Drachme » → (dirham = درهم), « Télévision » → (tilfaz = تلفاز). Ce procédé est l'un des facteurs les plus importants qui ont contribué à la modernisation rapide de la langue arabe et l'assimilation de vocabulaire d'origine étrangère.

- **La composition** (نحت) est une notion largement utilisée dans les études linguistiques descriptives pour désigner une unité linguistique composée d'éléments qui fonctionnent indépendamment l'un de l'autre dans d'autres circonstances. Il s'agit de la fusion de deux mots en un seul pour désigner un nouveau concept comme dans « آفوكاتو » (Avocat) : «Avo » = « عفو » (pardon) + « Cato » = « قاضي » (Cadi ou Juge) [celui qui cherche le "pardon" du "Juge"].

    La composition n'est pas un phénomène propre à la langue arabe, mais s'opère également dans d'autres langues comme le français ou l'anglais : « Brunch » (*Breakfeast + Lunch*), « télématique », « électromagnétique » …

Nous devons signaler toutefois, que chacune de ces méthodes a ses propres caractéristiques et ses propres limites.

Il est évident que la dérivation a joué un grand rôle dans la production d'une terminologie arabe depuis l'époque abbaside. C'est aussi la méthode la plus naturelle et la plus pratique aujourd'hui du fait que son application répond le mieux aux exigences du développement technologique récent qui transforme le monde arabe. La dérivation accélère certainement le transfert de nouveaux concepts en arabe.

Cependant, la dérivation peut aussi être vue comme une option peu appropriée en raison du manque de coordination entre néologues et universitaires arabes qui a engendré une surabondance de termes synonymes. La dérivation peut aussi être considérée de ce point de vue comme une technique peu efficace puisque la nature de la langue arabe impose des règles fixes et des formes syntaxiques qui ne peuvent pas être modifiés ou ignorés facilement pour couvrir le flot de synonymes. La langue arabe bien que considérée comme une langue de dérivation, n'accepte pas certaines formes nouvelles de mots qui ne suivent pas le modèle de la racine trilatérale.

Le procédé de l'arabisation a également servi de méthode très pratique pour créer des néologismes arabes depuis le début du XIXe siècle, quand le rôle de l'arabe comme langue de communication du savoir a commencé à décliner. L'arabisation est plus efficace dans le traitement de nouveaux termes techniques et scientifiques que la dérivation ou la composition. Elle peut mieux traiter un seul morphème en lui appliquant certaines règles grammaticales alors que dans un morphème composé, il est difficile de lui appliquer les mêmes règles. À titre d'exemple, on ne pourrait pas obtenir un équivalent uniterme arabe pour exprimer les notions de « sociolinguistique » (ألسنية اجتماعية [littéralement : « linguistique sociale »]), « intercontinental » (عبر القارات [littéralement : à travers les continents]), etc.

La méthode de composition, quant à elle, peut jouer un rôle intéressant dans l'affixation des termes étrangers et servir ainsi de mécanisme d'abréviation des termes arabes. Les termes français « immoral » (préfixe « im »), « décentralisation » (préfixe « dé »), ou le terme anglais « Wireless » (suffixe « less »), sont préfixés dans leurs équivalents arabes par l'article de négation « لا » :

« لاسلكي » ; « لامركزي » ; « لاأخلاقي ». Or, certaines académies arabes, particulièrement celle du Caire, ont émis des restrictions à cette forme de composition par affixe, pour eux étrangère à la langue arabe. De ce fait, ils considèrent qu'elle ne devrait être appliquée que dans les cas de nécessité scientifique absolue.

Néanmoins, le processus de composition est globalement moins productif que les deux autres méthodes car il n'a pas été décrit par les grammairiens arabes comme une méthode habituelle pour former de nouveaux termes. Il n'y a pas eu de déclaration directe de la part des philologues arabes quant à l'acceptabilité de la méthode de composition comme un procédé d'enrichissement de la langue arabe.

Force est de constater que la plupart des érudits et linguistes arabes considèrent les méthodes classiques traditionnelles de production néologique comme un legs intangible à l'abri de toutes forme de modification, soit en raison du lien « sacré » de la langue au texte du Coran, soit du fait des liens intellectuels très forts qui rattachent certains linguistes arabes contemporains aux pères fondateurs de la grammaire arabe de l'époque médiévale. Or, la langue n'est pas seulement un emblème du patrimoine spirituel ou culturel ; elle est aussi un outil et une production sociale dont la croissance et l'évolution sont analogues à la croissance et à l'évolution d'un peuple ou d'une nation. « La terminologie, tout en véhiculant les concepts organisés textuellement dans un domaine donné au sein d'une société, ne saurait trop s'éloigner de celle-ci, ni de ses idéologies, enjeux et fonctions qui se prêtent à une légitime discussion » (Rey 1985). De ce fait, nous croyons que l'une des premières recommandations à faire aux Académies arabes, serait d'enquêter sur les moyens de revitaliser les règles trop statiques de la terminologie afin d'offrir une plus grande flexibilité à la langue arabe pour qu'elle devienne plus pratique et plus fonctionnelle. Notre première recommandation dans ce sens, serait de suivre la voie de la normalisation internationale, car le problème fondamental de la terminologie arabe contemporaine est accentué par l'absence d'une normalisation fédératrice et créative qui pourrait (et qui devrait) faire face à la variété langagière dans le monde arabe (arabe classique, arabe moderne standard, arabe vernaculaire). Ces variétés ont rendu le lien entre le terme étranger et son équivalent arabe très confus. Le lien sémantique est aussi parfois difficile à situer puisque la relation entre le terme étranger, auquel les Arabophones sont déjà exposés, et son équivalent arabe n'est pas de prime abord une chose évidente.

Or, la normalisation de la terminologie et de la lexicographie est désormais un enjeu fondamental qui devient de plus en plus crucial à l'ère de la mondialisation multilingue, et plus stratégique encore en terminotique et en traductique.

La mise en place d'une normalisation coordonnée de la terminologie arabe au niveau de l'espace transnational arabophone aurait pour effet de réduire le temps de gestion linguistique, par exemple, pour construire et mettre à jour très rapidement des bases de données terminologiques dans plusieurs domaines de connaissance. Elle aura aussi pour effet de réduire le temps de localisation logicielle, puisque les équipes de traduction ne perdront plus de temps et de moyens à chercher les termes nécessaires pour chaque communauté d'usage. Le fait d'adopter des normes terminologiques communes augmentera certainement la qualité globale des produits industriels, scientifiques et culturels, et renforcera par la même occasion la diffusion et l'utilisabilité des produits sur le marché arabe. La normalisation de la terminologie favorisera surtout la cohérence linguistique entre les différentes régions du monde arabe et renforcera les consensus au sein de la communauté scientifique et technique arabophone. En résumé, les normes de terminologie augmenteront la qualité des produits et réduiront la marge de divergence entre les formes langagières de l'arabe (classique et vernaculaire). Enfin, il est certain que si le territoire arabophone ne se mobilise pour produire lui-même cet environnement terminologique normalisé, il est certain que des industriels des grands pays développés le lui rendront disponible et incontournable de l'extérieur ce qui n'est souhaitable ni sur le plan économique, ni scientifique, ni culturel.

En traductique, la normalisation terminologique est l'un des éléments essentiels pour qu'une communication adéquate soit assurée dans les langues cibles. La cohérence dans la relation entre le signifiant et le signifié est essentielle pour permettre une traduction correcte. Il existe, cependant, plusieurs exemples (en traduction) dans lesquelles une confusion apparaît entre la variation stylistique et l'incohérence dans l'usage de lexèmes. La variation stylistique est un moyen littéraire bien connu pour éviter les répétitions dans les textes en faisant appel aux synonymes. L'incohérence s'installe lorsqu'un signifié qui a été utilisé dans la langue cible pour rendre le sens d'un nouveau concept emprunté est utilisée en alternative inadéquate avec d'autres de ces synonymes. Le traducteur peut créer une certaine confusion lorsqu'il utilise un synonyme pour signifier le même concept plutôt que le lexème consacré. Dès lors le lecteur peut ne plus être en mesure de suivre la progression du texte, car il suppose qu'il existe une signification différente pour chaque synonyme.

En communication, la clé pour comprendre tout élément d'information est de connaître la terminologie qu'il comporte. La terminologie est un composé essentiel du transfert de l'information. Aujourd'hui, l'anglais se maintient comme la langue de science et de technologie dans le monde entier alors que l'arabe, qui avait été la langue de la science et de la technologie pendant le Moyen-âge, peine aujourd'hui à tenir sa position dans la diffusion de la civilisation et surtout des sciences et des technologies modernes. L'objectif principal pour les chercheurs terminologues contemporains arabes serait de déterminer l'applicabilité de certaines méthodes proposées par les premiers grammairiens arabes permettant de créer et d'introduire de nouveaux termes arabes pour faire face aux terminologies modernes qui augmentent continûment. Plusieurs méthodes sont disponibles: dérivation, arabisation, composition, translitération pour trouver des équivalents arabes de termes étrangers. Les résultats ont montré que chacune des méthodes pourrait jouer son rôle dans la résolution de la pénétration inévitable de termes étrangers qui sont créés presque tous les jours, en créant et en introduisant les homologues arabes les plus appropriés de ces termes (El Mgrab, 2011).

Quelles sont les voies de la normalisation moderne que nous pourrons proposer pour une terminologie arabe qui doit retrouver son lustre d'antan comme vecteur transmetteur des sciences, des techniques et de la diversité des cultures ? Comment envisager un processus palliatif à une sclérose linguistique arabe figée depuis le IXe siècle par de faux principes de sacralité religieuse et de nostalgies liturgiques de l'époque glorieuse de la poésie arabe préislamique des « Mu'allaqat » (les suspendues[19]) et des œuvres littéraires des premiers siècles de l'Islam ?

De par notre expérience dans le domaine normatif de la terminologie dans le domaine de l'e-Learning, nous ne saurions recommander autre moyen que celui des travaux du comité technique 37 de l'ISO dont les acquis normatifs ont été prouvés à plusieurs reprises depuis sa création en 1947.

---

[19] Sept poèmes lyriques du premier siècle avant l'Islam, encore très célèbres aujourd'hui dans le monde arabe pour l'excellence de leur style poétique et leur force langagière. Ils ont été écrits en caractères dorés et suspendus sur les murs de la « Ka'ba » (Mecque) pendant les foires annuelles de la poésie arabe de l'époque qui se déroulaient autour du sanctuaire religieux de la Mecque.

## 7. Les travaux de l'ISO/TC37

Nous n'aborderons pas ici les caractéristiques du TC37 en détail, d'une part en raison de l'actif normatif abondant de ce comité technique qui nécessite un espace plus important pour le présenter convenablement, et d'autre part pour avoir abordé ces travaux dans d'autres publications auxquels nous renvoyons.

Nous dirions toutefois que le TC37 de l'ISO est un « comité horizontal », qui fournit des lignes directrices pour tous les comités techniques qui élaborent des normes sur la façon de gérer leurs problèmes terminologiques. Les normes élaborées par l'ISO/TC37 ne sont pas toutefois limitées aux normes ISO. Elles cherchent aussi à s'assurer que les exigences et les besoins de tous les utilisateurs potentiels de normes concernant la terminologie, les contenus structurés et les langues, soient dûment traitées.

### *7.1. Un cadre de consensus*

Les normes de l'ISO/TC37 sont donc fondamentales et devraient constituer la base pour les solutions de localisation, de traduction et de toute autre forme d'application industrielle. Chez les praticiens de la normalisation, il y a un réel consensus général pour que la définition normative des concepts précède la normalisation des produits et des services, d'où leur utilité incontournable dans la sphère économique et industrielle. C'est pour cela qu'elles sont élaborées par des experts de l'industrie, des milieux universitaires et des entreprises qui sont délégués dans les instances de normalisation tant nationales qu'internationales. La participation est d'ailleurs ouverte à toutes les parties prenantes (*stakeholders*) qui œuvrent tous à l'élaboration de larges consensus entre les instituts nationaux de normalisation qui collaborent par voie d'adhésion.

Depuis sa création en 1947, l'ISO/TC37 dispose d'une longue histoire d'activités d'unification de la terminologie. Dans le passé, les experts en terminologie ont du se battre pour se faire connaitre. Aujourd'hui, leur expertise est sollicitée dans de nombreux domaines de la normalisation. La société de l'information multilingue et la société de la connaissance dépendra du contenu numérique fiable. La terminologie est ici indispensable. C'est parce que la terminologie joue un rôle crucial où et quand les informations et les connaissances spécialisées sont requises (par exemple dans la recherche et le développement), utilisées (par exemple dans les textes spécialisés), enregistrées et traitées (par exemple dans les banques de données), transmises (par la formation et l'enseignement), mises en œuvre (par exemple dans la technologie et le transfert de connaissances), ou traduites et interprétées (France-Russie Facilities, 2011).

À l'ère de la mondialisation, le besoin de normes méthodologiques concernant le contenu numérique multilingue est crucial. L'ISO/TC37 a développé au fil des ans, une expertise des normes méthodologiques au profit de la diversité linguistique numérique. Aujourd'hui, l'impact de ses travaux se concrétise autour de deux formes d'activités : une normalisation des terminologies et une normalisation des principes et des méthodes de constructions terminologiques. Les deux orientations sont mutuellement interdépendantes car la normalisation des terminologies n'entraînerait pas de données de haute qualité terminologiques si certains principes communs, des règles et des méthodes n'étaient pas appliqués. Ces principes, règles et méthodes terminologiques normalisées doivent aussi refléter l'état de l'art de développement de la théorie et de la méthodologie dans les domaines où les données terminologiques doivent être normalisées.

### *7.2. Profusion et exhaustivité d'un référentiel normatif*

Le TC37, marquera l'histoire de la terminologie normalisée dans tous les domaines par une production abondante et exhaustive de normes terminologiques grâce à sa structuration en sous comités spécialisés chargés des Principes et méthodes (TC 37/SC : 1) ; des Méthodes de travail terminographiques et lexicographiques (TC 37/SC 2) ; des systèmes de gestion de la terminologie, de la connaissance et du contenu (TC 37/SC 3) ; de la gestion des ressources linguistiques (TC 37/SC 4) ; de la traduction, interprétation et des technologies apparentées (TC 37/SC 5). Ces sous comités ont dans leur actif une série de normes qui constitue un référentiel incontournable pour tous le travail de normalisation dans les comités et les sous comités de l'ISO, du CEI (Commission Électrotechnique Internationale) et des autres instances de normalisation comme le CEN, l'IEEE, le W3C, etc. Parmi les normes phares du TC37 :

- ISO 704 : Travail terminologique – Principes et méthodes
- ISO 860 : Travaux terminologiques – Harmonisation des concepts et des termes
- ISO 10241-1 : Articles terminologiques dans les normes – Partie 1 : Exigences générales et exemples de présentation
- ISO 16642 : Plate-forme pour le balisage de terminologies informatisées
- ISO 30042 : TermBase eXchange (TBX)
- ISO 29383 : Politiques terminologiques – Élaboration et mise en œuvre
- ISO 22128 : Produits et services en terminologie – Aperçu et orientation
- ISO 12616: Terminographie axée sur la traduction
- ISO 12620 : Registre de catégories de données

Toutes ces normes se valent en efficacité, tellement elles ont été réétudiées et révisées selon les directives de l'ISO[20] pour répondre en permanence aux besoins du marché d'une part et aux spécificités culturelles et linguistiques des communautés de pratique de l'autre.

On constate à travers cette série de normes, que le TC37 a commencé par définir son champ d'application, ses principes et méthodes dans la norme 704. Cette norme définit les principes fondamentaux et les méthodes de la préparation et de la compilation des terminologies, qu'il s'agisse de créer des terminologies, des vocabulaires, des nomenclatures pour l'usage strict des experts en normalisation ou pour des visées plus larges que ce soit dans un cadre industriel ou linguistique.

Le TC37 a normalisé ensuite un catalogue ouvert de catégories de données apte à définir des données terminologiques ou lexicographiques (ISO/IEC 12620), puis un cadre commun de mise en œuvre terminotique à même d'assurer l'interopérabilité et la réusabilité des ressources terminologiques indépendamment des diverses banques de données terminologiques. Ce cadre commun, le TMF (*Terminological Markup Framework* : ISO 16642)[21] que nous avons très souvent décrit dans d'autres publications, nécessite bien sûr que ces différentes bases respectent le métamodèle XML TMF, ou exige que les ressources terminologiques soient reformatées selon ce même modèle. Le TMF détermine les principes généraux qui contribuent à la gestion de la formation des désignations et à la formulation des définitions. Ce modèle décrit les liens entre les objets, les concepts et leurs représentations par des terminologies, une approche qui a suscité beaucoup de débats autour de deux méthodes constitutives des données terminologiques : sémasiologique et onomasiologique.

---

[20] Mise à jour systématique de toute norme publiée tous les 5 ans.
[21] Un cadre de mise en œuvre normalisé pour la terminotique, lui-même lié au TML (*Terminological Markup Language*), un langage XML spécialisé (en fait une DTD) pour l'expression des données terminologique.

### *7.3. Le débat de méthodes : entre approche sémasiologique et approche onomasiologique*

Par méthode sémasiologique, les rédacteurs monolingues, les traducteurs, et dans une certaine mesure les lexicographes (notamment les rédacteurs de dictionnaires de langue ou des dictionnaires bilingues), entendent partir des termes, des mots ou des expressions (syntagmes) pour pointer sur leur signification (des concepts et leur définition).

Par méthode onomasiologique, les documentalistes, les spécialistes du *knowledge management*, les ingénieurs ou les chercheurs analysant des processus techniques ou un phénomène scientifique, entendent avoir une démarche de direction opposée. Il faut partir des concepts, que l'on organise en système de concepts (avec des graphes de relations : génériques, partitives, relationnelles), pour pointer, sur un deuxième niveau, sur des jargons ou des langues de spécialité scientifiques ou techniques (et ce, dans autant de langues – multilinguisme - que nécessaire). Sans vouloir lui dérober son utilité, la pratique onomasiologique constitue le point de départ où ancrer la complémentarité éventuelle des deux perspectives onomasiologiques et sémasiologiques du rapport terme/concept en terminologie » (Aito, Igwe, 2011).

Pour ce qui est de l'aspect humain de l'activité terminologique ou lexicographique, la synergie entre les deux directions de méthode rapproche étroitement les zones frontalières entre la sémasiologie, réservée jusqu'ici en terminologie à la lexicologie, et l'onomasiologie, approche qui convient à la terminologie traditionnelle wüsterienne. Toutes les terminologies se réalisent en liant les deux méthodes en synergie dialectique. Aujourd'hui la communauté des terminologues, mais aussi celle des lexicographes, se sont entendus pour normaliser une démarche unique. Celle-ci est onomasiologique. Au bout d'un long processus de discussion, la communauté des normalisateurs terminologues et lexicographes ont fini par adopter comme seule méthode valide, la démarche onomasiologique (ISO704, Terminologie : principes et méthodes).

## 8. **Pour conclure : entreprendre la numérisation structurée et normalisée de très grands corpus en langue arabe est un enjeu majeur pour la traductique du futur**

Comment présenter désormais aux institutions et aux chercheurs arabophones l'urgence des mesures à entreprendre dans la dynamique normative, multilingue et multiécriture qui est en train de changer le monde de la recherche scientifique ? Quelles sont surtout les actions clés sur lesquelles fonder une stratégie de repositionnement de la langue et de l'écriture arabe parmi les langues et les écritures les plus utilisées au monde ? La voie de la normalisation terminologique est certes une alternative *sine qua non*, pour les raisons que nous avons évoquées précédemment quant aux défaillances dans l'harmonisation des procédés terminologiques déployés, et la pauvreté des néologismes arabes dans les domaines des sciences et des techniques. Mais, nous supposons que d'autres solutions sont aussi importantes et peuvent constituer des catalyseurs réels vers une unification terminologique et conceptuelle de la langue arabe. C'est par la mise en ligne de très grands corpus de ressources linguistiques arabes (mais aussi surtout bilingues), non pas uniquement de ressources textuelles numérisées, mais aussi de corpus oraux ou de collections numérisées de cinéma, de TV, de radio, de disques, etc. que la dynamique entre pratiques de constitution de terminologies et praxis de leur usage, voire leur marchandisation éventuelle, s'établira. Le volume et la diversité des ressources numériques arabes sur les réseaux sont le tremplin par lequel tous les autres aspects culturels et linguistiques seront réformés et optimisés par des normes appropriées. Il existe, en effet, de plus en plus de normes soit sémantiques (comme la TEI, *Text Encoding Initiative*), soit morphosyntaxiques pour la description normalisée des ressources linguistiques (comme celle de l'ISO TC37), normes de description des caractéristiques de trait qui devient un atout et un enjeu primordial pour une traductique en devenir utilisant comme référent et comme contexte l'accès à de gigantesques corpus de ressources linguistiques mondialisées. La langue et l'écriture arabe doivent s'inscrire absolument dans ces chantiers. À l'ère de la mondialisation des réseaux et des systèmes d'information numériques multilingues, ce n'est plus un choix, c'est une question de survie culturelle et identitaire.